\RequirePackage{lineno}
\documentclass[twocolumn,showpacs,preprintnumbers,linenumbers,superscriptaddress,amsmath,amssymb]{revtex4}
\usepackage{graphicx}
\usepackage{longtable}
\usepackage{color}
\usepackage{cases}

\begin{document}
\title{Probing Parton Dynamics of QCD Matter with $\Omega$ and $\phi$ Production}

\affiliation{AGH University of Science and Technology, Cracow 30-059, Poland}
\affiliation{Argonne National Laboratory, Argonne, Illinois 60439, USA}
\affiliation{Brookhaven National Laboratory, Upton, New York 11973, USA}
\affiliation{University of California, Berkeley, California 94720, USA}
\affiliation{University of California, Davis, California 95616, USA}
\affiliation{University of California, Los Angeles, California 90095, USA}
\affiliation{Central China Normal University (HZNU), Wuhan 430079, China}
\affiliation{University of Illinois at Chicago, Chicago, Illinois 60607, USA}
\affiliation{Creighton University, Omaha, Nebraska 68178, USA}
\affiliation{Czech Technical University in Prague, FNSPE, Prague, 115 19, Czech Republic}
\affiliation{Nuclear Physics Institute AS CR, 250 68 \v{R}e\v{z}/Prague, Czech Republic}
\affiliation{Frankfurt Institute for Advanced Studies FIAS, Frankfurt 60438, Germany}
\affiliation{Institute of Physics, Bhubaneswar 751005, India}
\affiliation{Indian Institute of Technology, Mumbai 400076, India}
\affiliation{Indiana University, Bloomington, Indiana 47408, USA}
\affiliation{Alikhanov Institute for Theoretical and Experimental Physics, Moscow 117218, Russia}
\affiliation{University of Jammu, Jammu 180001, India}
\affiliation{Joint Institute for Nuclear Research, Dubna, 141 980, Russia}
\affiliation{Kent State University, Kent, Ohio 44242, USA}
\affiliation{University of Kentucky, Lexington, Kentucky, 40506-0055, USA}
\affiliation{Korea Institute of Science and Technology Information, Daejeon 305-701, Korea}
\affiliation{Institute of Modern Physics, Lanzhou 730000, China}
\affiliation{Lawrence Berkeley National Laboratory, Berkeley, California 94720, USA}
\affiliation{Max-Planck-Institut fur Physik, Munich 80805, Germany}
\affiliation{Michigan State University, East Lansing, Michigan 48824, USA}
\affiliation{Moscow Engineering Physics Institute, Moscow 115409, Russia}
\affiliation{National Institute of Science Education and Research, Bhubaneswar 751005, India}
\affiliation{Ohio State University, Columbus, Ohio 43210, USA}
\affiliation{Institute of Nuclear Physics PAN, Cracow 31-342, Poland}
\affiliation{Panjab University, Chandigarh 160014, India}
\affiliation{Pennsylvania State University, University Park, Pennsylvania 16802, USA}
\affiliation{Institute of High Energy Physics, Protvino 142281, Russia}
\affiliation{Purdue University, West Lafayette, Indiana 47907, USA}
\affiliation{Pusan National University, Pusan 609735, Republic of Korea}
\affiliation{University of Rajasthan, Jaipur 302004, India}
\affiliation{Rice University, Houston, Texas 77251, USA}
\affiliation{University of Science and Technology of China, Hefei 230026, China}
\affiliation{Shandong University, Jinan, Shandong 250100, China}
\affiliation{Shanghai Institute of Applied Physics, Shanghai 201800, China}
\affiliation{Temple University, Philadelphia, Pennsylvania 19122, USA}
\affiliation{Texas A\&M University, College Station, Texas 77843, USA}
\affiliation{University of Texas, Austin, Texas 78712, USA}
\affiliation{University of Houston, Houston, Texas 77204, USA}
\affiliation{Tsinghua University, Beijing 100084, China}
\affiliation{United States Naval Academy, Annapolis, Maryland, 21402, USA}
\affiliation{Valparaiso University, Valparaiso, Indiana 46383, USA}
\affiliation{Variable Energy Cyclotron Centre, Kolkata 700064, India}
\affiliation{Warsaw University of Technology, Warsaw 00-661, Poland}
\affiliation{Wayne State University, Detroit, Michigan 48201, USA}
\affiliation{World Laboratory for Cosmology and Particle Physics (WLCAPP), Cairo 11571, Egypt}
\affiliation{Yale University, New Haven, Connecticut 06520, USA}
\affiliation{University of Zagreb, Zagreb, HR-10002, Croatia}

\author{L.~Adamczyk}\affiliation{AGH University of Science and Technology, Cracow 30-059, Poland}
\author{J.~K.~Adkins}\affiliation{University of Kentucky, Lexington, Kentucky, 40506-0055, USA}
\author{G.~Agakishiev}\affiliation{Joint Institute for Nuclear Research, Dubna, 141 980, Russia}
\author{M.~M.~Aggarwal}\affiliation{Panjab University, Chandigarh 160014, India}
\author{Z.~Ahammed}\affiliation{Variable Energy Cyclotron Centre, Kolkata 700064, India}
\author{I.~Alekseev}\affiliation{Alikhanov Institute for Theoretical and Experimental Physics, Moscow 117218, Russia}
\author{J.~Alford}\affiliation{Kent State University, Kent, Ohio 44242, USA}
\author{A.~Aparin}\affiliation{Joint Institute for Nuclear Research, Dubna, 141 980, Russia}
\author{D.~Arkhipkin}\affiliation{Brookhaven National Laboratory, Upton, New York 11973, USA}
\author{E.~C.~Aschenauer}\affiliation{Brookhaven National Laboratory, Upton, New York 11973, USA}
% ...

\collaboration{STAR Collaboration}\noaffiliation

\begin{abstract}
We present measurements of $\Omega$ and $\phi$ production at mid-rapidity from Au+Au collisions at
nucleon-nucleon center-of-mass energies $\sqrt{s_{\tiny{\textrm{NN}}}} = 7.7$, $11.5$, $19.6$, $27$ and $39$ GeV
by the STAR experiment at the Relativistic Heavy Ion Collider (RHIC). Motivated by the coalescence formation mechanism for these strange hadrons, we study the ratios of
\emph{N}($\Omega^{-}+\overline{\Omega}^{+}$)/(2\emph{N}($\phi$)). These ratios as a function of transverse momentum ($p_T$) fall on a consistent trend at high collision energies, but start to show deviations in peripheral collisions at $\sqrt{s_{\tiny{\textrm{NN}}}} = 19.6$, 27 and 39 GeV, and in central collisions at 11.5 GeV in the intermediate $p_{T}$ region of 2.4$-$3.6 GeV/$c$. We further evaluate empirically the strange quark $p_T$ distributions at hadronization by studying the $\Omega/\phi$ ratios scaled by the number of constituent quarks. The NCQ-scaled $\Omega/\phi$ ratios show a suppression of strange quark production in central collisions
at 11.5 GeV compared to $\sqrt{s_{\tiny{\textrm{NN}}}} \geq 19.6$ GeV. The shapes of the presumably
thermal strange quark distributions in 0-60\% most central collisions at 7.7 GeV show significant deviations from those in 0-10\% most central collisions at higher energies.
These features suggest that there is likely a change of the underlying strange quark dynamics in the transition from quark-matter to hadronic matter at collision energies below 19.6 GeV.

\end{abstract}



\maketitle


Lattice quantum chromodynamics (QCD) calculations suggest that, at high temperature and low
baryon chemical potential ($\mu_{\textrm{B}}$), the transition from
the Quark Gluon Plasma (QGP) to the state of a hadron gas is
smooth and continuous (cross-over transition) \cite{crossover}. At
lower temperatures and high $\mu_{\textrm{B}}$, theoretical
calculations predict a first order phase transition \cite{1storder}
which may end at a critical point \cite{cpoint}.
The mapping of the QCD phase diagram has been a subject
of intensive theoretical and experimental activities in the past decades.
In central Pb-Pb collisions at Super Proton Synchrotron (SPS),
the enhanced production of $\Omega$ at $\sqrt{s_{\tiny{\textrm{NN}}}}$ = 8.8
and 17.3 GeV \cite{na49st, na49ome, na57ome, na57ome1} and
$\phi$ mesons at $\sqrt{s_{\tiny{\textrm{NN}}}}$ = 6.3$-$17.3 GeV \cite{na49phi}
compared to $\pi$ mesons has been considered as a QGP signal \cite{raf}.
Multi-strange hadrons such as $\Omega(sss)$ hyperons and
$\phi(s\overline{s})$ mesons are important probes for the search of the QCD phase boundaries
\cite{philong, phincqb}.
The $\Omega$ hyperons and $\phi$ mesons are
expected to have relatively small hadronic interaction cross
sections \cite{ashor, nxuplot}. Therefore, they can carry the
information directly from the chemical freeze-out stage with little or no
distortion due to hadronic rescattering. In addition, the measured $\Omega$ and $\phi$ yields
suffer minimal distortion from decay feed-down.
As a result, the production of the $\Omega$ and $\phi$ particles offers a
unique advantage in probing the transition from partonic to hadronic
dynamics.

In heavy ion collisions at the top RHIC energy of $\sqrt{s_{\tiny{\textrm{NN}}}}$ = 200 GeV,
model calculations~\cite{reconbination0, reconbination1,
reconbination2, reconbination3, jinhui1} and experimental data
suggest that particles at intermediate $p_T$ are formed via the coalescence of
low $p_{T}$ quarks from the bulk partonic matter and/or fragmented hard partons.
Experimentally, baryon to meson ratios have been found to be large compared to
those from elementary collisions~\cite{phenixpip, starpip, philong, starcucu, aliceksla2013}. The measured
elliptic flow $v_2$ has been found to scale with the
number of constituent quarks (NCQ) for both baryons and mesons \cite{STAR-Lambda-v2}
in Au+Au collisions at the top RHIC energy.
In order to explain these observations, coalescence model calculations require
the development of collectivity among constituent quarks during the partonic phase.
This partonic collectivity has been considered as an important evidence for the formation of
deconfined QCD matter with partonic degrees of freedom in Au+Au
collisions at the highest RHIC energy~\cite{phenixpip,
starpip,philong, starcucu, aliceksla2013}.

In order to map out the phase diagram of the QCD
matter, a Beam Energy Scan (BES) program has been initiated at RHIC
with Au+Au collisions at $\sqrt{s_{\tiny{\textrm{NN}}}}=7.7-39$ GeV~\cite{starbes}. These collisions allow us to
reach a broad range of temperature and
$\mu_{B}$ in the QCD phase diagram \cite{tmub} and search for a possible beam energy region where the underlying
dynamics are different from those of partonic matter observed in Au+Au collisions at the top RHIC
energy.

The STAR experiment, which has a large acceptance detector system \cite{Ackermann:2002ad},
has collected Au+Au collision data at
$\sqrt{s_{\tiny{\textrm{NN}}}}=7.7$, 11.5 and 39 GeV in 2010, and
at $\sqrt{s_{\tiny{\textrm{NN}}}} = 19.6$ and 27 GeV in 2011.
Compared to previous measurements at SPS \cite{na49st, na49ome, na57ome, na57ome1, na49phi},
RHIC offers wide range of beam energies. In addition, the STAR detector provides uniform acceptance over different energies
and extensive reach to the intermediate $p_{T}$ range for both $\Omega$ and $\phi$ mesons for different centralities.
In this letter, we present the first STAR measurements of mid-rapidity ($|y|<0.5$)
$\Omega$ and $\phi$ production with a broad $p_T$ coverage for various collision centrality bins at selected energies for the BES.
With guidance from coalescence formation mechanism for primordial particles such as $\Omega$ and $\phi$, we examine features in particle production and
explore possible changes from bulk partonic coalescence to hadron dominated dynamics as the colliding energy decreases.

A minimum bias trigger was used to record Au+Au collision events in this analysis.
The trigger was defined using a coincidence of signals from either the Zero Degree Calorimeters,
Vertex Position Detectors or Beam-Beam Counters
\cite{beschgv2, bespidlongv2}. STAR's Time Projection Chamber (TPC) \cite{tpc} was
used for tracking of charged particles and particle identification. In the offline data
analysis, we require the radial position of the reconstructed primary
vertex to be within 2 cm of the beam axis to suppress events
from collisions with the beam pipe (radius of 3.95 cm).
To ensure nearly uniform detector acceptance, the analyzed events were
required to have a primary $Z$ vertex (along beam direction) within
$\pm70$ cm from the
center of the TPC for $\sqrt{s_{\tiny{\textrm{NN}}}} = 7.7$, $19.6$, and $27$ GeV collisions and $\pm50$ cm, $\pm40$ cm for $\sqrt{s_{\tiny{\textrm{NN}}}} = 11.5$ and $39$ GeV,
respectively. After the event selection, we obtain approximately 4, 12, 36,
70 and 130 million Au+Au minimum bias triggered events at
$\sqrt{s_{\tiny{\textrm{NN}}}} = 7.7$, 11.5, 19.6, 27 and 39 GeV, respectively. The
collision centrality was determined by comparing the uncorrected
charged hadron multiplicity measured from the TPC at mid-rapidity ($|\eta|<$ 0.5) with Monte Carlo Glauber simulations
\cite{beschgv2, bespidlongv2}.

The multi-strange hadron signals and raw yields were obtained from
the invariant mass distributions reconstructed through their hadronic
decay channels: $\phi\rightarrow K^{+}+K^{-}$ and
$\Omega^{-}(\overline{\Omega}^{+})\rightarrow \Lambda(\bar{\Lambda}) +
K^{-}(K^{+})$. The decay daughters $\Lambda(\bar{\Lambda})$ were
reconstructed through $\Lambda(\bar{\Lambda})\rightarrow
p(\bar{p})+\pi^{-}(\pi^{+})$. Charged hadrons
($\pi^{\pm}, K^{\pm}, p, \bar{p}$) were identified by their specific
energy loss (d$E$/d$x$) in the TPC gas \cite{tpc}. The
combinatorial background of the weakly decaying particles
$\Lambda(\bar{\Lambda})$ and $\Omega^{-}(\overline{\Omega}^{+})$ was reduced by
geometrical cuts on their decay topology~\cite{omerecon, bespidlongv2}. The $\Omega^{-}(\overline{\Omega}^{+})$ combinatorial
background was estimated by rotating $K^{-}$($K^{+}$) tracks at 5 different angles from $\pi/3$ to $5\pi/3$
and normalizing the invariant mass distribution to the mass window of (1.625 GeV/$c^{2}$, 1.655 GeV/$c^{2}$) and
(1.69 GeV/$c^{2}$, 1.72 GeV/$c^{2}$).
The $\Omega^{-}$($\overline{\Omega}^{+}$) raw yields were extracted by counting the signals
within a mass window from 1.660 to 1.685 GeV/$c^{2}$ after subtracting the rotational background.
The $K^{+}K^{-}$ combinatorial background in $\phi$ meson reconstruction was
subtracted with the mixed event technique \cite{philong, plbphi}.
The $\phi$ meson raw yields were determined by a Breit-Wigner + polynomial function
(up to second order) fit to the mixed-event-background-subtracted $K^{+}K^{-}$ invariant mass
distribution \cite{philong, plbphi}.

\begin{figure}[htbp]
\centering \vspace{0.1cm}
\includegraphics[width=8.5cm]{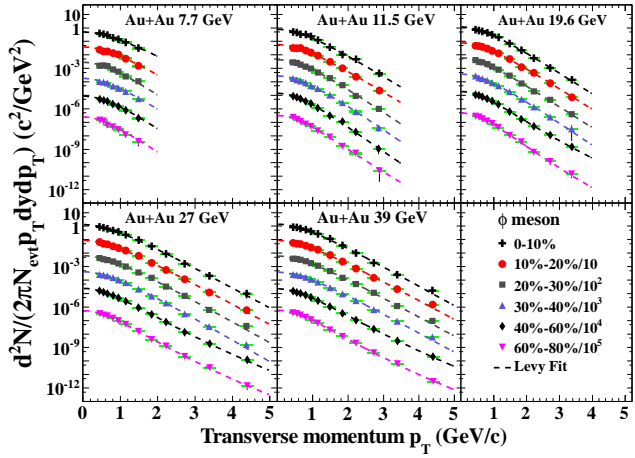}
\vspace{0cm} \caption{\label{fig1.fig} (Color online)
Mid-rapidity ($|y|<0.5$) $\phi$ meson $p_{T}$ spectra from
Au+Au collisions at different centralities and energies
($\sqrt{s_{\tiny{\textrm{NN}}}} = 7.7$ - 39 GeV). The green bands represent
systematic errors. The dashed curves represent fits to the
experimental data with a Levy function \cite{philong}.}
\end{figure}
\begin{figure}[htbp]
\centering
\includegraphics[width=8.7cm]{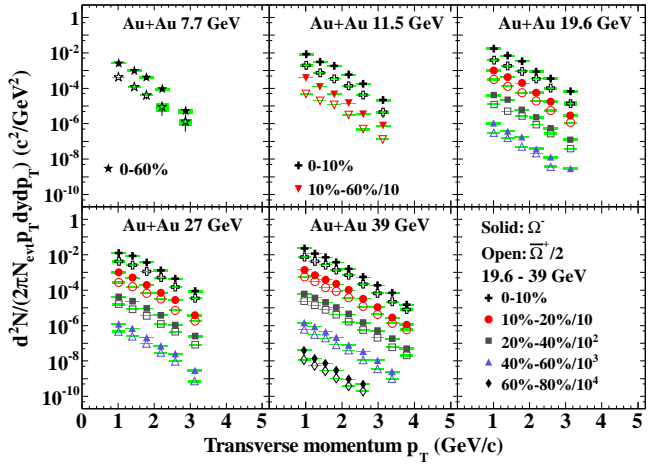}
\vspace{-4ex}\caption{\label{fig2.fig} (Color online) Mid-rapidity
($|y|<0.5$) $\Omega^{-}$($\overline{\Omega}^{+}$) $p_{T}$ spectra from
Au+Au collisions at different centralities and energies
($\sqrt{s_{\tiny{\textrm{NN}}}} = 7.7$ - 39 GeV).
The open symbols represent $\overline{\Omega}^{+}$ and solid symbols represent $\Omega^{-}$.
The green bands denote systematic errors.}
\end{figure}

Figures 1 and 2 show the $p_{T}$ spectra of $\phi$, and
$\Omega^{-}(\overline{\Omega}^{+})$ at mid-rapidity ($|y| < 0.5$) for different
centralities from Au+Au collisions at
$\sqrt{s_{\tiny{\textrm{NN}}}} = 7.7 - 39$ GeV. The spectra were corrected for reconstruction efficiency and geometrical
acceptance. The systematic errors mainly come from two sources: the different signal extraction techniques,
and the reconstruction efficiency corrections.
They were studied as a function of $p_T$ and were obtained by exploring the dependence of invariant yields on various raw yield extraction
techniques including different fit/counting ranges and different fit functions,
and on different combinations of analysis cuts.
For the $\phi$ meson, relative systematic errors of invariant yields vary from 10\%-16\% at $\sqrt{s_{\tiny{\textrm{NN}}}} =
11.5-39$ GeV to 17\%-21\% at $\sqrt{s_{\tiny{\textrm{NN}}}} =
7.7$ GeV. The systematic errors in 0-10\% central collisions are generally larger than those
in 60\%-80\% peripheral collisions by 2\%-3\% due to greater combinatorial backgrounds.
For $p_{T} < 0.8$ GeV/$c$ in central collisions, the uncertainty of $\phi$ meson
raw yield extraction is dominant. However, for $p_{T}>1.6$ GeV/$c$
the main source of systematic error is the differences in track selection cuts.
For the $\Omega$ invariant yields, the relative systematic errors vary from $\sim5$\% to 20\%,
and are dominated by the signal extraction methods. Due to the higher combinatorial background
in $p_{T}\lesssim1.2$ GeV/$c$ and low statistics at $p_{T}\gtrsim2.8$ GeV/$c$,
the systematic errors are found to be larger in the corresponding $p_{T}$ windows.
The systematic uncertainties have a weak centrality dependence and their
energy dependences for $\Omega$ and $\phi$ particles are similar.
The systematic errors of invariant yields of $\phi$ and $\Omega$ are shown
as green bands in Figs. 1 and 2 for each $p_{T}$ bin.

\begin{figure}[htbp]
\centering
\includegraphics[width=7.7cm]{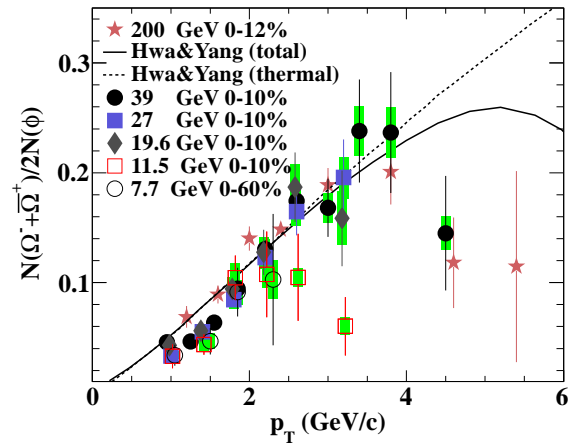}
\vspace{-2ex} \caption{(Color online) The baryon-to-meson ratio,
\emph{N}($\Omega^{-}+\overline{\Omega}^{+}$)/(2\emph{N}($\phi$)), as a function
of $p_{T}$ in mid-rapidity ($|y| < 0.5$) from Au+Au
collisions at $\sqrt{s_{\tiny{\textrm{NN}}}} = 7.7\ -$ 200 GeV.
The green bands denote systematical errors. The solid and dashed lines
represent recombination model calculations for central collisions at
$\sqrt{s_{\tiny{\textrm{NN}}}} = 200$ GeV \cite{reconbination1} with
total and thermal strange quark contributions, respectively.
\label{fig3.fig}}
\end{figure}

\begin{figure}[t]
\vspace{0ex}\centering
\includegraphics[width=8cm]{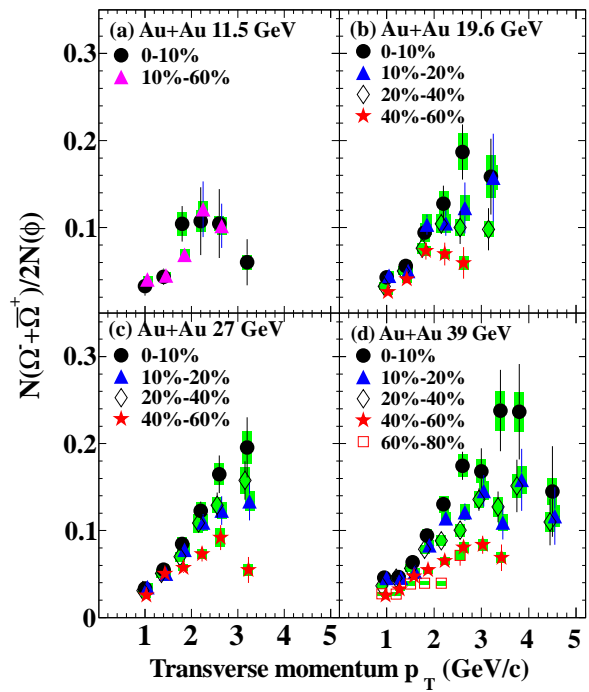}
\vspace{-2ex} \caption{\label{fig4.fig}(Color online) Centrality
dependence of \emph{N}($\Omega^{-}+\overline{\Omega}^{+}$)/(2\emph{N}($\phi$))
ratios, as a function of $p_{T}$ in mid-rapidity ($|y| < 0.5$) from
Au+Au collisions at $\sqrt{s_{\tiny{\textrm{NN}}}} = 11.5$, 19.6, 27 and
39 GeV. The green bands denote systematical errors.}
\end{figure}

We present baryon-to-meson ratios,
\emph{N}($\Omega^{-}+\overline{\Omega}^{+}$)/(2\emph{N}($\phi$)), as a function
of $p_{T}$ from Au+Au collisions for various beam energies from
$\sqrt{s_{\tiny{\textrm{NN}}}} = 7.7$ to $200$ GeV in Fig.~\ref{fig3.fig} and for various collision centralities in Fig.~\ref{fig4.fig}, respectively. Data from 200 GeV Au+Au collisions
are from previously published STAR results \cite{philong}. Coalescence or recombination models~\cite{reconbination1,
reconbination2, reconbination3} have been used to describe particle productions in nucleus-nucleus collisions at RHIC. In particular, a model calculation by Hwa and Yang for Au+Au collisions at $\sqrt{s_{\tiny{\textrm{NN}}}} = 200$
GeV~\cite{reconbination1} predicted that most of the $\Omega$ and $\phi$ yields up to the intermediate $p_T$ region are from coalescence/recombination of thermal strange quarks. The straight dotted line assumed that these thermal strange quarks have exponential $p_T$ distributions. Deviations from the straight line at high $p_T$ were attributed to recombination with strange quarks from high $p_T$ showers. Deviations from the theory calculation at low $p_T$ could indicate that thermal strange quarks may not have an exponential distribution.
Possibly, other particle production dynamics may also contribute deviations from coalescence model calculations.

In Fig.~\ref{fig3.fig} the measured \emph{N}($\Omega^{-}+\overline{\Omega}^{+}$)/(2\emph{N}($\phi$)) ratios from central
Au+Au collisions at $\sqrt{s_{\tiny{\textrm{NN}}}} = 19.6$, 27 and
39 GeV follow closely the ratio from 200 GeV and are consistent with a picture of coalescence/recombination dynamics over
a broad $p_T$ range of $1-4$ GeV/\emph{c}. The ratios at 11.5
GeV seem to deviate from the trend observed at higher beam energies.
In particular, the ratios at 11.5 GeV appear to turn down
around $p_T$ of 2 GeV/$c$ while those at higher beam energies such as 39 and 200 GeV
peak at $p_T$ of 3 GeV/$c$ or above. The collision centrality dependence of the
\emph{N}($\Omega^{-}+\overline{\Omega}^{+}$)/(2\emph{N}($\phi$)) ratios in Fig. \ref{fig4.fig}(a)-(d) shows
distinct differences between the $40\%-60\%$ centrality bin and the other
centrality bins for Au+Au collisions at 19.6 and 27 GeV. Furthermore, the ratios from the
peripheral collisions of $40\%-60\%$ at 27 GeV are similar in magnitude to the
ratios from collisions at 11.5 GeV. Because the $\Omega$ and $\phi$ particles have small hadronic rescattering
cross sections~\cite{radflow}, the change in these ratios is likely to originate from the partonic phase. The decrease in the \emph{N}($\Omega^{-}+\overline{\Omega}^{+}$)/(2\emph{N}($\phi$))
ratios from central collisions at 11.5 GeV compared to those at 19.6 GeV or above may indicate a significant change
in the hadron formation dynamics and/or in strange quark $p_{T}$ distributions at the lower energy.
Such a change may arise from a transition from hadronic to partonic dynamics with increasing
beam energy. The turn-over in the ratios from Au+Au collisions below 11.5 GeV beam energy is unlikely due to contributions of high $p_T$ shower partons as suggested by model calculation from Hwa and Yang \cite{reconbination1} because of  relatively low $p_T$ particles involved.

We illustrate qualitatively the change in the underlying bulk strange quark $p_T$ distribution by following the
procedure developed in Ref. \cite{jinhui1}. We assume that the $\Omega$ baryons are formed from coalescence of three strange quarks of approximately equal momentum
and the $\phi$ mesons from two strange quarks. In the coalescence framework, the
$\Omega$ baryon production probability is proportional to the local strange quark density, $f_{s}^{3}(p_{T}^{s})$,
 and the $\phi$ meson is proportional to $f_{s}(p_{T}^{s})f_{\overline{s}}(p_{T}^{\overline{s}})$, where $f_{s}$($f_{\overline{s}}$) is the strange (anti-strange) quark $p_{T}$ distribution at hadronization.
Assuming that strange quarks and anti-strange quarks have a similar
slope of $p_{T}$ distribution, the NCQ-scaled ratio
$\frac{N(\Omega^{-}+\overline{\Omega}^{+})|_{p_{T}^{\Omega}=3p_{T}^{s}}}{2N(\phi)|_{p_{T}^{\phi}=2p_{T}^{s}}}$
could reflect the strange quark distribution at hadronization.
We note that theoretical calculations with a more sophisticated
recombination scheme have been developed by He \emph{et al.}
\cite{he} and the extracted strange quark
distribution is similar to that from the present approach.

\begin{figure}[ht]
\vspace{-1ex}\centering
\includegraphics[width=7.5 cm]{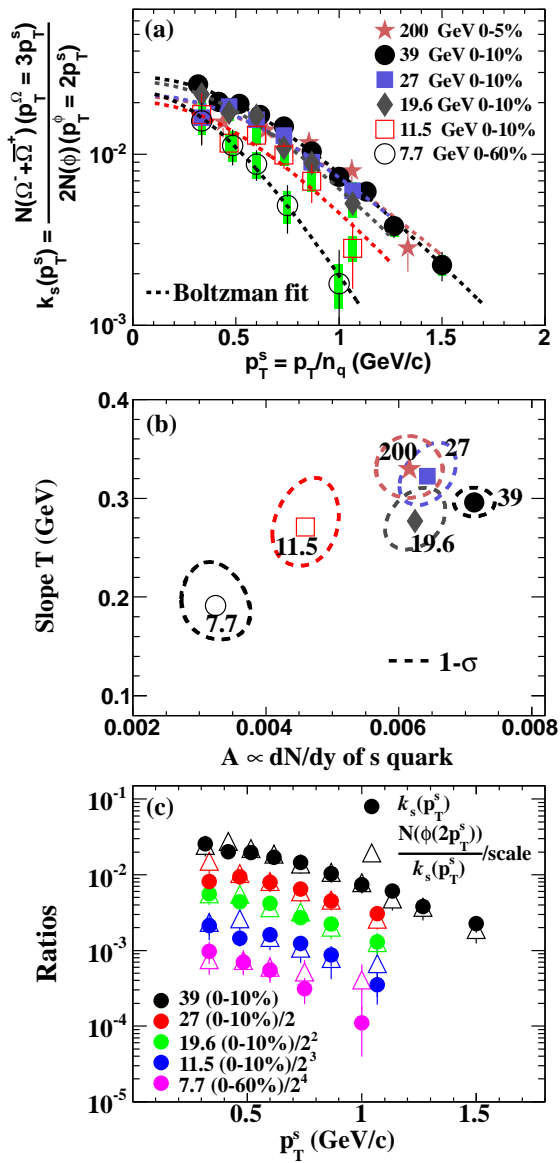}
\vspace{-1ex} \caption{\label{fig5.fig}(Color online) (a) NCQ-scaled
\emph{N}($\Omega^{-}+\overline{\Omega}^{+}$)/(2\emph{N}($\phi$)) ratios, as a
function of $p_{T}/n_{q}$ in mid-rapidity ($|y| < 0.5$) from
Au+Au collisions at $\sqrt{s_{\tiny{\textrm{NN}}}} = 7.7 - 200$
GeV. Here $n_{q}$ is the number of constituent quarks of each
hadron. The green bands denote systematic errors. Dashed lines are
Boltzmann fits to data.
(b) The fitting parameters $A$ and $T$, and $1\sigma$ contours (including statistic and systematic errors).
(c) The ratios of $f_{s}(p_{T}^{s})$ and scaled $\frac{N(\phi(2p_{T}^{s}))}{f_{s}(p_{T}^{s})}$ as a function of $p_{T}^{s}$.
The scale factors are 394.4, 763.7, 742.8, 870.9, and 746.5 for $\sqrt{s_{\tiny{\textrm{NN}}}} = 7.7$, 11.5, 19.6, 27, and 39 GeV, respectively.
}
\end{figure}

Figure \ref{fig5.fig}(a) shows the NCQ-scaled
$\frac{N(\Omega^{-}+\overline{\Omega}^{+})|_{p_{T}^{\Omega}=3p_{T}^{s}}}{2N(\phi)|_{p_{T}^{\phi}=2p_{T}^{s}}}$
ratios as a function of $p_{T}^{s} = p_{T}/n_{q}$ at mid-rapidity
($|y| < 0.5$) from central Au+Au collisions at
$\sqrt{s_{\tiny{\textrm{NN}}}} = 11.5 - 200$ GeV as well as 0-60\% collisions at 7.7 GeV.
Here $N(\Omega^{-}+\overline{\Omega}^{+})$ and $N(\phi)$ denote invariant yields
of $(\Omega^{-}+\overline{\Omega}^{+})$ and $\phi$, respectively. Since the $p_{T}$ bin widths used for the
$\Omega^{-}(\overline{\Omega}^{+})$ and $\phi$ meson spectra do not match, we use our Levy fit (see Fig. \ref{fig1.fig}) to interpolate the invariant yield of $\phi$ meson at desired $p_T$. The NCQ-scaled
$\frac{N(\Omega^{-}+\overline{\Omega}^{+})|_{p_{T}^{\Omega}=3p_{T}^{s}}}{2N(\phi)|_{p_{T}^{\phi}=2p_{T}^{s}}}$
ratios at all energies can be fit
with a Boltzmann distribution $\frac{g_{s}Am_{T}}{T(m_{s}+T)}e^{-(m_{T}-m_{s})/T}$,
where $m_s$ is the effective strange quark mass of 0.46 GeV/$c^2$ from
Ref.~\cite{reconbination2}, $m_T$ is the transverse mass ($\sqrt{m_s^2+p_T^2}$), $T$ is
the slope parameter of the exponential function which may be related to
the freeze-out temperature and radial expansion velocity of strange quarks \cite{omerecon}.
Considering different yields ratios of $\overline{s}$ quark over $s$ quark with collision energies,
that is, $f_{s}(p_{T}^{s})=r(\sqrt{s_{\tiny{\textrm{NN}}}})f_{\overline{s}}(p_{T}^{\overline{s}})$, where
$r^{3}(\sqrt{s_{\tiny{\textrm{NN}}}})=\frac{dN}{dy}(\overline{\Omega}^{+})/\frac{dN}{dy}(\Omega^{-})$,
we include a correction factor $g_{s}=(1+r^{3})/r$ in the Boltzmann distribution function (based on the coalescence calculation \cite{reconbination1}), and then $A$ is proportional to strange quark rapidity density.

The fitting parameters $A$ and $T$, and $1\sigma$ contours are shown in Fig. \ref{fig5.fig}(b).
Figs. \ref{fig5.fig}(a)-(b) show that the derived strange quark distributions
vary little in shape as a function of beam energy from 11.5 GeV to 200 GeV.
The amplitude parameter $A$ at 11.5 GeV, however,
seems to be noticeably smaller than those data of 19.6 GeV or above. Based on coalescence model \cite{reconbination1}, the smaller strange quark local density at 11.5 GeV is probably responsible for the smaller \emph{N}($\Omega^{-}+\overline{\Omega}^{+}$)/(2\emph{N}($\phi$)) ratios as shown in Fig. 3, where the first two low $p_{T}$ points at 11.5 GeV are systematically lower than those at $\sqrt{s_{\tiny{\textrm{NN}}}} \geq 19.6$ GeV. At 7.7 GeV, the slope parameter $T$ is smaller than those data of 19.6 GeV or above,
with a 1.8$\sigma$ standard deviation from the 19.6 GeV result. We note that one
possible reason for the deviation of $T$ is the centrality difference
since the data at 7.7 GeV are for 0-60\% while those at other energies are for central collisions.
In the framework of the coalescence mechanism, our derived ratio distribution
can be sensitive to both the local density and the $p_{T}$ distribution of strange quarks.
Our data of 19.6 GeV or above show little beam energy dependence suggesting strange quark
equilibration may have been approximately achieved in those central collisions,
possibly due to strange quark dynamics rather than hadronic processes \cite{koch86}.
The variation of the 11.5 GeV data may arise from the strangeness non-equilibration
and the presence of a strangeness phase space suppression factor ($\gamma_s < 1$) \cite{omerecon}.
A possible transition in the collision dynamics and in the dominate
degrees of freedom (partonic versus hadronic)
below 19.6 GeV needs further experimental investigation
with more experimental probes and with larger data samples \cite{bespidv2}.

Recently, the ALICE experiment reported an observation of nearly flat ratios of proton to $\phi$ as a function of $p_T$ in central Pb+Pb collisions at $\sqrt{s_{\tiny{\textrm{NN}}}} = 2.76$ TeV \cite{aliceksla}.
It was argued that the similarity in shapes of the $p_T$ spectra indicates that the radial flow of these particles is mostly determined by the masses of
these particles as hydrodynamical calculations predicted, instead of by the number of constituent quarks as expected from coalescence models~\cite{aliceksla}. The ALICE proton to $\phi$ ratios as a function of $p_T$ also showed a very strong dependence on collision centrality. We note that such a strong dependence is an indication that the protons undergo considerable rescattering during the hadronic evolution as expected from the hybrid model calculation~\cite{radflow}. It is important to disentangle the hadronic rescattering contributions to the radial flow for ordinary hadrons in order to address the partonic flow prior to hadronization. We have used the $\Omega$ and $\phi$ spectra to carry out another independent check on the coalescence picture: we can obtain the strange quark $p_T$ distribution from the quark number scaled $\Omega$ to $\phi$ ratios as in Fig.~\ref{fig5.fig}(a); and we can calculate another strange quark $p_T$ distribution by dividing the $\phi$ with the previously obtained strange quark distribution. Fig.~\ref{fig5.fig}(c) shows these two strange quark distributions. The consistency of these distributions indicates that there is one unique strange quark distribution which can explain both $\Omega$ and $\phi$ $p_T$ spectra, a necessary condition for coalescence model.

In summary, STAR has measured the production of multi-strange
hadrons $\Omega$ and $\phi$ at mid-rapidity from Au+Au collisions at
$\sqrt{s_{\tiny{\textrm{NN}}}} = 7.7$, 11.5, 19.6, 27 and 39 GeV
from the BES program at RHIC.
The \emph{N}($\Omega^{-}+\overline{\Omega}^{+}$)/(2\emph{N}($\phi$)) ratios at
intermediate $p_{T}$ in peripheral collisions are found to be
lower than those in central collisions at 19.6, 27 and 39 GeV. The ratios from
11.5 GeV central collisions are systematically lower than those from collisions at 19.6 GeV or above
for $p_{T}>2.4$ GeV/$c$. The NCQ-scaled $\Omega/\phi$ ratios show a suppression of strange quark production
in 11.5 GeV compared to $\sqrt{s_{\tiny{\textrm{NN}}}} \geq 19.6$ GeV. The shapes of the presumably
thermal strange quark distributions in 0-60\% most central collisions at 7.7 GeV show significant deviations from those in 0-10\% most central collisions at higher energies.
These features suggest that
there is likely a change in the underlying strange quark dynamics
in the bulk QCD matter responsible for $\Omega$
and $\phi$ production. Our measurements point to collision energies below 19.6 GeV for further investigation of a possible transition
from partonic dominant matter ($\sqrt{s_{\tiny{\textrm{NN}}}}>19.6$ GeV) to hadronic dominant matter ($\sqrt{s_{\tiny{\textrm{NN}}}} < 11.5$ GeV).



We thank the RHIC Operations Group and RCF at BNL,
the NERSC Center at LBNL, the KISTI Center in Korea,
and the Open Science Grid consortium for providing resources and support. This work was
supported in part by the Office of Nuclear Physics within the U.S. DOE Office of Science,
the U.S. NSF, the Ministry of Education and Science of the Russian Federation, NNSFC, CAS,
MoST (973 Program No. 2014CB845400) and MoE of China, the Korean Research Foundation, GA and MSMT of the Czech Republic,
FIAS of Germany, DAE, DST, and UGC of India, the National Science Centre of Poland, National
Research Foundation, the Ministry of Science, Education and Sports of the Republic of Croatia,
and RosAtom of Russia. We thank Rudolph C. Hwa for valuable discussions.

\end{document}